\documentclass[11pt,letterpaper]{article}

\usepackage[utf8]{inputenc}
\usepackage[T1]{fontenc}
\usepackage[margin=1in]{geometry}
\usepackage{amsmath,amssymb,amsthm,mathtools}
\usepackage{graphicx}
\usepackage{booktabs}
\usepackage{natbib}
\usepackage{setspace}
\usepackage{float}
\usepackage{caption}
\usepackage{enumitem}
\usepackage[dvipsnames,table]{xcolor}
\usepackage{tikz}
\usetikzlibrary{positioning,arrows.meta,fit,backgrounds}
\usepackage[colorlinks=true,linkcolor=blue!60!black,citecolor=blue!60!black,urlcolor=blue!60!black,breaklinks=true]{hyperref}

\newcommand{\Prob}{\mathbb{P}}
\newcommand{\TE}{\operatorname{TE}}
\newcommand{\code}[1]{\texttt{#1}}

\title{\textbf{Econstellar: An Open-Source AI-Augmented Research Engine\\
for Computational Financial Econometrics}}

\author{
Avishek Bhandari\thanks{School of Humanities, Social Sciences and Management, Indian Institute of Technology Bhubaneswar. Email: \texttt{avishekb@iitbbs.ac.in}. Code and live demonstration: \url{https://avishekb9.github.io/econstellar/}.}
}

\date{\today}

\begin{document}
\maketitle

\begin{abstract}
\noindent Turning a promising economic idea into a credible empirical finding is, in
practice, an expensive undertaking: it demands a great deal of specialised computation, and
the results are seldom released in a form that others can check or build upon. Econstellar
is our response. It is an open, publicly served research engine that runs publication-grade
financial econometrics from an ordinary web browser and explains what the results mean, so
that a reader does not merely read a finding but can re-run it, vary its inputs, and trace
exactly how it was produced. Three choices give the system its character. The heavy
computation is placed on the processor that suits it, rather than forced onto hardware
ill-matched to the task, which is much of the reason analysis of this kind is so rarely
served to the public. An artificial-intelligence assistant selects and interprets the
analyses but never originates a number, so every quantity it reports is a real computation
the reader can reproduce. And the engine a visitor exercises is the same code that produced
the figures in our published research. We expose seventeen econometric methods, each
reported with a verified live value and reproducible at the public endpoint, computed under
a single discipline: prices are treated as non-stationary and all methods are applied to
returns. The system also regenerates, on demand, the headline result of an accompanying
study of financial contagion, from the package that generated it. The platform is the
working core of an active research programme spanning three software releases and three
preprints, and it is available now, free and open-source, at a live public address. Our aim
is a simple one: to shorten the distance between a research claim and the moment another
person can independently verify it.
\end{abstract}



\smallskip
\noindent\textbf{ACM CCS:} Software and its engineering $\rightarrow$ Software
architectures; Applied computing $\rightarrow$ Economics; Information systems
$\rightarrow$ Web applications.

\onehalfspacing

\section{Introduction}\label{sec:intro}

The distance between a suggestive working paper and a defensible empirical submission in
computational economics is, more often than not, one to two orders of magnitude of
floating-point work, paid almost entirely in compute. A network-econometric result that
is gestured at in a seminar, an estimated spillover or a directed information-flow graph,
becomes a publishable object only after surrogate nulls, bootstrap intervals, rolling
re-estimation, and robustness across specifications have been computed, and that arithmetic
is rarely exposed in a form a reader can re-execute. It may be noted that the replication
literature has documented the consequence directly: a substantial fraction of published
empirical economics cannot be reproduced from the materials provided
\citep{ChangLi2022}, and a parallel literature in the computational sciences has codified
the practices that would prevent it \citep{Sandve2013}. The gap is not principally one of
intent; it is that the computation itself is heavy, irregular, and seldom served.

A natural response would be to accelerate the heavy step on graphics processors, as has
been done with conspicuous success for dynamic-equilibrium models whose inner loops are
dense and regular \citep{Aldrich2011}. The central information-theoretic primitive of
network contagion analysis, however, resists that route. Directed information flow between
return series is measured by transfer entropy \citep{Schreiber2000}, and its exact
estimator in continuous state spaces is the nearest-neighbour construction of
\citet{KraskovStoegbauerGrassberger2004}, which rests on $k$-d-tree range searches. The
control flow of a $k$-d-tree traversal is data-dependent and branch-divergent: queries
issued together descend different subtrees, prune at different nodes, and terminate at
different depths. On single-instruction-multiple-thread accelerators such divergence
serialises the threads of a warp, because the hardware must execute the taken and untaken
sides of each branch in turn \citep{Fung2007}. The computation is, in consequence,
pointer-chasing and memory-latency-bound rather than arithmetic-bound, which is precisely
the profile for which graphics processors are ill-suited. Exact information-theoretic
contagion estimation is, in short, a central-processing-unit workload, and the
architectural premise of the system described here follows from that fact: the heavy
econometrics is sited on the processor that suits it, and only the interpretive and
data-gathering layers are delegated to managed services.

What does not exist, to our knowledge, is an open, sandboxed, publicly served econometrics
engine that combines information-theoretic contagion estimation of this kind with
artificial-intelligence-augmented interpretation and live reproducibility, so that a reader
may not only read a result but
re-run it, vary its parameters, and obtain a grounded explanation, all from a browser and
without installing anything. Econstellar is built to close that gap. It is a research
engine, in the literal sense that the same deployed code that produces the figures of the
group's substantive papers is the code a visitor exercises at the public endpoint. The
present paper documents the system: its architecture and the security model that a public
computation endpoint requires (Section~\ref{sec:architecture}); the methodological
substrate of eight primitives that the seventeen served methods instantiate
(Section~\ref{sec:substrate}); the seventeen methods with their verified live values and
the stationarity discipline under which they are computed (Section~\ref{sec:results}); and
the published software and preprints that the engine supports, together with the
reproduce-the-paper facility that ties them to live computation
(Section~\ref{sec:outputs}); and its live surfaces and how to navigate them
(Section~\ref{sec:availability}). Section~\ref{sec:conclude} concludes.

\section{Architecture}\label{sec:architecture}

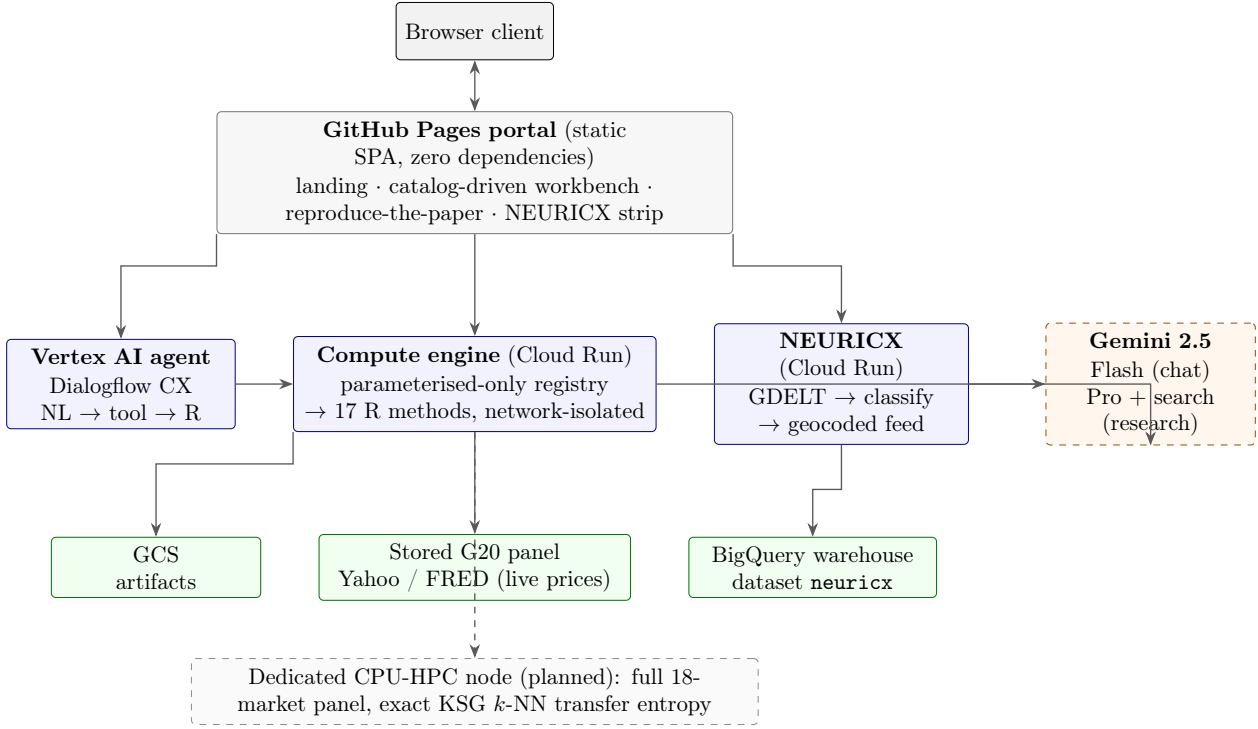
\begin{figure}[!tbp]
\centering
\resizebox{\textwidth}{!}{%
\begin{tikzpicture}[
  font=\small,
  >={Stealth[length=2.2mm]},
  node distance=8mm and 9mm,
  box/.style={draw, rounded corners=2pt, align=center, inner sep=4pt, minimum height=9mm},
  cli/.style={box, fill=black!5},
  spa/.style={box, fill=black!3, draw=black!55},
  svc/.style={box, fill=blue!5, draw=blue!55!black},
  ext/.style={box, fill=orange!7, draw=orange!60!black, dashed},
  store/.style={box, fill=green!6, draw=green!45!black},
  plan/.style={box, fill=black!2, draw=black!45, dashed},
  flow/.style={->, semithick, draw=black!65},
  bi/.style={<->, semithick, draw=black!65},
]
\node[cli] (client) {Browser client};
\node[spa, below=of client, text width=78mm] (portal)
  {\textbf{GitHub Pages portal} (static SPA, zero dependencies)\\[1pt]
   landing $\cdot$ catalog-driven workbench $\cdot$ reproduce-the-paper $\cdot$ NEURICX strip};
\node[svc, below=16mm of portal, text width=54mm] (engine)
  {\textbf{Compute engine} (Cloud Run)\\[1pt]
   parameterised-only registry\\ $\rightarrow$ 17 R methods, network-isolated};
\node[svc, left=of engine, text width=33mm] (agent)
  {\textbf{Vertex AI agent}\\[1pt] Dialogflow CX\\ NL $\rightarrow$ tool $\rightarrow$ R};
\node[svc, right=of engine, text width=37mm] (neuricx)
  {\textbf{NEURICX} (Cloud Run)\\[1pt] GDELT $\rightarrow$ classify\\ $\rightarrow$ geocoded feed};
\node[ext, right=12mm of neuricx, text width=30mm] (gemini)
  {\textbf{Gemini 2.5}\\[1pt] Flash (chat)\\ Pro $+$ search (research)};
\node[store, below=16mm of engine, text width=46mm] (data)
  {Stored G20 panel\\ Yahoo / FRED (live prices)};
\node[store, left=of data, text width=30mm] (gcs) {GCS\\ artifacts};
\node[store, right=of data, text width=36mm] (bq)
  {BigQuery warehouse\\ dataset \texttt{neuricx}};
\node[plan, below=9mm of data, text width=86mm] (hpc)
  {Dedicated CPU-HPC node (planned): full 18-market panel, exact KSG $k$-NN transfer entropy};
\draw[bi] (client) -- (portal);
\draw[flow] (portal) -- (engine);
\draw[flow] (portal.south west) -- ++(0,-0.5) -| (agent.north);
\draw[flow] (portal.south east) -- ++(0,-0.5) -| (neuricx.north);
\draw[flow] (agent) -- (engine);
\draw[flow] (neuricx) -- (gemini);
\draw[flow] (engine.east) -| (gemini.south);
\draw[flow] (engine) -- (data);
\draw[flow] (engine.south west) -- ++(0,-0.5) -| (gcs.north);
\draw[flow] (neuricx.south) -- ++(0,-0.7) -| (bq.north);
\draw[flow, dashed] (engine) -- (hpc);
\end{tikzpicture}%
}
\caption{The Econstellar system. A static single-page application on GitHub Pages is the
only public surface a reader touches; it reads the engine's method catalogue at runtime
and renders results client-side. Three Cloud Run services sit behind it: the sandboxed
compute engine, which exposes seventeen registered econometric methods over a
parameterised-only registry and executes each in a network-isolated R subprocess; the
NEURICX news-intelligence service; and a Vertex AI conversational agent. The language
model (Gemini 2.5) is called only to interpret results and classify news, never to
originate quantities. Persistent state lives in Google Cloud Storage and a BigQuery
longitudinal warehouse. A dedicated CPU-HPC node (dashed) is the planned route to the
full eighteen-market panel and an exact nearest-neighbour transfer-entropy
implementation.}
\label{fig:architecture}
\end{figure}

The system has four parts: a sandboxed compute engine, a two-tier analyst built on a
language model, a live news-intelligence service with a longitudinal warehouse, and a
public workbench. The reader touches only the last of these. Figure~\ref{fig:architecture}
gives the overall arrangement: a static single-page application on GitHub Pages is the sole
public surface, and it reads the engine's method catalogue at runtime and renders results
client-side, while three Cloud Run services and two storage back-ends sit behind it. We
take the four parts in turn.

\subsection{The sandboxed compute engine}\label{sec:engine}

The compute engine is a zero-dependency Node.js orchestrator that exposes a single
registered analysis as an HTTP endpoint. The public surface accepts a method name and a
set of typed parameters, validates them against a fixed registry schema, and never accepts
executable code. An accepted call spawns one R runner script in a disposable subprocess
with no network egress, a read-only root, an ephemeral scratch directory, and a wall-clock
timeout; an unrecognised method name or a malformed parameter is rejected before any
process is spawned. The primary security guarantee is therefore the registry itself: the
endpoint can perform exactly the finite, reviewed list of registered analyses and nothing
besides. On the local development host the subprocess is additionally confined by a user
namespace with the network unshared (\code{bwrap --unshare-net}); on Cloud Run the same
network isolation and wall-clock timeout are enforced within the platform's managed
container sandbox, and a live run reports its sandbox state as network-isolated with a
timeout rather than as a privileged in-process jail, which is the honest description of
what the managed environment grants.

It is pertinent to describe the threat model explicitly, because the endpoint is public by
design and the design is intentional. The adversary is any anonymous caller; the asset to
be protected is the underlying compute account and its credentials; and the attack surface
is the request body. Because the registry admits only a method name and typed parameters,
there is no path by which a caller can cause arbitrary code to execute, exfiltrate data
over the network from inside a runner, or escalate beyond the finite menu of analyses. The
remaining risk is resource exhaustion, which is bounded by layered limits: per-address rate
limits of twenty, ten, and five requests per minute for the compute, fast-analyst, and
deep-research endpoints respectively; per-address daily ceilings on the two paid analyst
endpoints; a global concurrency ceiling that sheds load rather than queueing it unboundedly;
and a global daily ceiling on calls to the language model, so that the monetary exposure of
the public endpoint is capped irrespective of the number of callers. We notice that these
limits are a property of the orchestrator, not of the network perimeter, and so they hold
uniformly across the local and deployed configurations.

\subsection{The artificial-intelligence analyst layer}\label{sec:ai}

Interpretation is provided in two tiers, and both are constructed so that the language
model chooses and explains analyses without ever producing their numbers. The fast analyst
answers a natural-language question by function calling: the model is given a single tool,
\code{run\_analysis(method, params)}, whose schema is generated from the same method
registry that governs the compute endpoint, so the model can request only registered
analyses with validated arguments, and it sees the engine's actual JSON output before it
composes a reply. The deep-research assistant runs in two phases. In the first, an agentic
loop executes live econometrics through the same tool, accumulating results; in the second,
those results are summarised into context and a grounded synthesis is produced with real
citations retrieved by web search. The two phases are deliberately not combined into a
single model request, because pairing search grounding with function calling in one call is
fragile; separating them keeps each phase simple and auditable.

The key property of this layer is that the language model does not originate numbers. Every
quantitative claim in a generated answer is the output of a sandboxed computation that the
reader can reproduce through the same public endpoint; the model's role is confined to
selecting which registered analysis to run, reading its structured result, and composing an
interpretation in prose. This is the central architectural decision of the analyst layer,
and it is what distinguishes a grounded research assistant from a fluent but unverifiable
one: the engine is the source of every figure, and the model is the source only of the
words around it. A third interface, a Vertex AI conversational agent, exposes the same
tool through a managed dialogue runtime, and has been confirmed end-to-end from a
natural-language request to an R computation, returning, for example, an augmented
Dickey-Fuller statistic of $-52.64$ for United Kingdom equity returns, which is the value
the compute endpoint returns for that series.

\subsection{NEURICX: live financial intelligence}\label{sec:neuricx}

The third service, NEURICX, supplies live financial-news intelligence. It draws articles
from the GDELT global news index, classifies each by transmission channel with a language
model, geocodes the result, and renders it as a map feed; when the upstream index rate-limits
the service, a disk cache supplies a stale fallback so that the feed degrades rather than
fails. Each classified pull is persisted to a BigQuery warehouse, so that channel intensity,
sentiment, and a systemic-stress index accumulate as a queryable longitudinal record rather
than a transient view. The warehouse is what turns a live feed into a research input: a
question about how a given channel's intensity evolved across an episode is answered from
stored history rather than re-scraped on demand.

\subsection{The public research workbench}\label{sec:workbench}

The workbench is a single-page application written in dependency-free JavaScript and served
as static files from GitHub Pages. It reads the engine's catalogue at runtime and never
hardcodes a method name, so the interface for a method appears as soon as the method is
registered: enumerated parameters render as menus, numeric parameters as typed inputs, and
each of the seventeen methods has a dedicated result renderer, including a force-directed
drawing of the estimated network from layout coordinates returned by the engine. A
deep-research pane and a news strip surface the other two services in the same page. Two
properties make the workbench a reproducibility instrument and not merely a console. First,
every result is stamped with its provenance: the method, its version, the engine revision
that produced it, the parameters, the data vintage, a timestamp, and a permalink that
re-runs the identical analysis. Second, results export as CSV or JSON and carry a generated
BibTeX entry, and a dedicated page regenerates the headline results of the group's
published work live, through the same engine that produced them, so that a reader can move
from a claim in a paper to its recomputation in a single step.

\section{The Methodological Substrate}\label{sec:substrate}

The seventeen served methods are not a miscellany; they instantiate a small substrate of
eight estimation primitives that recur across the group's substantive work on contagion,
long memory, and network formation. We state each primitive with its defining equation, in
the form used in the group's methodological work, and then map the served methods onto them
in Table~\ref{tab:substrate}. The equations are reproduced for reference and completeness;
the contribution of this section is the observation that a compact and reusable substrate
underlies an apparently diverse method menu.

\paragraph{P1: long memory.} Detrended fluctuation analysis measures the scaling of the
integrated, locally detrended return profile and returns the Hurst exponent $H$
\citep{Peng1994}:
\begin{equation}\label{eq:dfa}
F(s) = \sqrt{\frac{1}{N}\sum_{i=1}^{N}\bigl(Y(i)-Y_{\mathrm{fit}}(i)\bigr)^2}\;\propto\; s^{H},
\end{equation}
where $Y(i)=\sum_{t=1}^{i}(x_t-\bar x)$ is the integrated return profile, $N$ the series
length, $Y_{\mathrm{fit}}$ the local least-squares polynomial trend within each window of
length $s$, and $H>\tfrac12$ signals persistent long memory.

\paragraph{P2: multi-scale variance.} The maximal-overlap discrete wavelet transform
decomposes return variance scale by scale \citep{PercivalWalden2000,GencaySelcukWhitcher2005}:
\begin{equation}\label{eq:modwt}
\sigma_x^2 = \sum_{j\ge 1}\nu_x^2(\tau_j),\qquad
\nu_x^2(\tau_j)=\frac{1}{N}\sum_{t} W_{j,t}^2,
\end{equation}
where $\nu_x^2(\tau_j)$ is the wavelet variance at scale $\tau_j=2^{\,j-1}$ and $W_{j,t}$
are the level-$j$ wavelet coefficients.

\paragraph{P3: directed information flow.} Transfer entropy measures directed dependence
\citep{Schreiber2000},
\begin{equation}\label{eq:te}
\TE_{Y\to X}=\sum p\bigl(x_{t+1},x_t^{(d)},y_t^{(d)}\bigr)
\log\frac{p\bigl(x_{t+1}\mid x_t^{(d)},y_t^{(d)}\bigr)}{p\bigl(x_{t+1}\mid x_t^{(d)}\bigr)},
\end{equation}
with $x_t^{(d)},y_t^{(d)}$ the delay embeddings. Its exact estimation in continuous spaces
uses the Kozachenko-Leonenko differential-entropy estimator,
\begin{equation}\label{eq:kl}
\widehat{H}(Z) = -\psi(k)+\psi(N)+\log c_m+\frac{m}{N}\sum_{i=1}^{N}\log\varepsilon_i,
\end{equation}
and the nearest-neighbour mutual-information correction of
\citet{KraskovStoegbauerGrassberger2004},
\begin{equation}\label{eq:ksg}
\widehat{I}(X;Y)=\psi(k)+\psi(N)-\frac{1}{N}\sum_{i=1}^{N}
\bigl[\psi(n_{x,i}+1)+\psi(n_{y,i}+1)\bigr],
\end{equation}
where $\psi$ is the digamma function, $\varepsilon_i$ twice the distance to the $k$th
neighbour, $c_m$ the volume of the unit $m$-ball with $m$ the embedding dimension, and
$n_{x,i},n_{y,i}$ the marginal neighbour counts; in the Gaussian case the
measure coincides with Granger causality \citep{BarnettBarrettSeth2009}. It is the
$k$-d-tree search underlying \eqref{eq:ksg} whose irregular control flow motivates the
processor choice of Section~\ref{sec:intro}.

\paragraph{P4: surrogate significance.} Directional significance is assessed against an
ensemble of phase-preserving surrogates \citep{TheilerEtAl1992}:
\begin{equation}\label{eq:surr}
\hat p_{Y\to X}=\frac{1+\#\{\,b:\TE^{(b)}_{\mathrm{surr}}\ge\widehat{\TE}_{Y\to X}\,\}}{B+1},
\end{equation}
the one-sided empirical $p$-value of the observed flow against $B$ surrogate replicates.

\paragraph{P5: community structure.} Community structure is summarised by modularity
\citep{Newman2006}:
\begin{equation}\label{eq:mod}
Q=\frac{1}{2m}\sum_{i,j}\Bigl(A_{ij}-\frac{k_ik_j}{2m}\Bigr)\delta(c_i,c_j),
\end{equation}
with $A$ the weighted adjacency, $k_i$ node strength, $m=\tfrac12\sum_i k_i$, and
$\delta(c_i,c_j)=1$ when nodes share a community.

\paragraph{P6: network formation.} The directed dependency graph retains only edges that
survive the surrogate test \citep{DieboldYilmaz2012}:
\begin{equation}\label{eq:edges}
E=\{(i\to j):\hat p_{i\to j}<\alpha\}.
\end{equation}

\paragraph{P7: channel attribution.} Measured contagion is attributed to economic channels
by an instrumented regression \citep{BhandariContagionChannels2026}:
\begin{equation}\label{eq:attr}
C_{ij}=\sum_{c}\beta_c\,Z^{(c)}_{ij}+u_{ij},
\end{equation}
where $Z^{(c)}_{ij}$ are channel proxies (trade, financial, and related) and $\beta_c$,
estimated by two-stage least squares with external instruments, is each channel's
transmission share.

\paragraph{P8: discriminatory validation.} Crisis-classification performance is measured by
the area under the receiver-operating-characteristic curve, with competing indices compared
by the correlated-AUC test of \citet{DeLong1988}:
\begin{equation}\label{eq:auc}
\mathrm{AUC}=\Prob\bigl(s_{\mathrm{crisis}}>s_{\mathrm{calm}}\bigr),
\end{equation}
the probability that the index scores a randomly chosen crisis period above a randomly
chosen calm period.

\begin{table}[!htbp]
\centering
\caption{The eight substrate primitives and the served methods that instantiate them.
A method may draw on more than one primitive; the dominant primitives are listed.}
\label{tab:substrate}
\small
\begin{tabular}{lll}
\toprule
Primitive & Equation & Served methods (Section~\ref{sec:results}) \\
\midrule
P1 long memory            & \eqref{eq:dfa}   & \code{dfa\_hurst} \\
P2 multi-scale variance   & \eqref{eq:modwt} & \code{wavelet}, \code{wavelet\_coherence}, \code{soch\_profile} \\
P3 directed information   & \eqref{eq:te}--\eqref{eq:ksg} & \code{wqte}, \code{granger}, \code{soch\_profile}, \code{quantile\_var} \\
P4 surrogate significance & \eqref{eq:surr}  & \code{wqte}, \code{soch\_profile} \\
P5 community structure     & \eqref{eq:mod}   & \code{network} \\
P6 network formation       & \eqref{eq:edges} & \code{granger}, \code{network}, \code{connectedness} \\
P7 channel attribution     & \eqref{eq:attr}  & \code{connectedness}, \code{spillover\_rolling} \\
P8 discriminatory validation & \eqref{eq:auc} & systemic-risk validation (Section~\ref{sec:results}) \\
\midrule
\multicolumn{3}{l}{\emph{Standard time-series machinery:} \code{unit\_root},
\code{live\_unit\_root}, \code{panel\_unit\_root}, \code{var\_irf}, \code{vecm},} \\
\multicolumn{3}{l}{\code{garch}, \code{rolling\_dcc} provide the stationarity gate,
dynamics, and volatility on which the primitives rest.} \\
\bottomrule
\end{tabular}
\end{table}

\section{Verified Results}\label{sec:results}

The seventeen methods are listed in Table~\ref{tab:methods}, each with a value produced by
the deployed engine. We wish to be explicit about the status of these numbers: every value
in the table was produced by the live sandboxed engine on the stored G20 equity panel and
is reproducible at the public endpoint by issuing the corresponding registered call. The
four most recently added methods were re-run, uncached, at the time of writing, and the
remainder were verified live against the deployed revision. The values are reported as the
engine returns them and are not rounded differently for presentation.

It is pertinent to state the stationarity discipline that governs the entire menu, since it
is the single methodological rule on which the validity of the others depends. Price levels
are integrated of order one, and returns are stationary; transfer entropy and every other
method in the table are therefore applied to log-returns throughout, and the engine never
reports a price level as stationary. The \code{live\_unit\_root} method makes the rule
operational on live data, testing levels and returns separately and reporting, for the
S\&P~500, a level series integrated of order one and a return series of order zero.

\begin{table}[!htbp]
\centering
\caption{The seventeen served methods with verified live values on the stored G20 equity
panel (eighteen markets, daily log-returns, 2006--2026). Each value is produced by the
deployed engine and is reproducible at the public endpoint. India, USA, and UK denote the
equity indices of those economies.}
\label{tab:methods}
\footnotesize
\begin{tabular}{lll}
\toprule
Method & Class & Verified live value \\
\midrule
\code{unit\_root}        & stationarity        & India ADF $-49.18$; UK $-52.64$ \\
\code{live\_unit\_root}  & stationarity (live) & S\&P levels $I(1)$ / returns $I(0)$ \\
\code{panel\_unit\_root} & stationarity (panel)& 5-market: IPS $-77.26$, LLC $-51.79$ \\
\code{var\_irf}          & dynamics            & lag 7, maximal root $0.705$ (stable) \\
\code{vecm}              & cointegration       & India/USA/UK rank 3 (trace $7265.97\to1851.19$) \\
\code{garch}             & volatility          & India $\alpha+\beta=0.991$ \\
\code{rolling\_dcc}      & time-varying corr.\ & India--USA mean $\rho=0.23$ ($0.15$--$0.31$) \\
\code{dfa\_hurst}        & long memory (P1)    & India $H=0.542$ \\
\code{wavelet}           & multi-scale (P2)    & India d1 $=47.07\%$ of variance \\
\code{wavelet\_coherence}& multi-scale (P2)    & USA/India mean $0.249$, peak d6 \\
\code{wqte}              & contagion (P3,P4)   & USA$\to$India $\tau_{.05}$ agg.\ $0.039$, rising d1--d4 \\
\code{soch\_profile}     & contagion (P2,P3)   & USA$\to$India agg.\ $0.039$, peak d4 (published package) \\
\code{quantile\_var}     & tail dynamics (P3)  & India/USA/UK $\tau_{.05}$: USA top driver (net $+0.70$) \\
\code{granger}           & causality (P3,P6)   & 6-market: 6 edges, USA out-degree 3 \\
\code{connectedness}     & spillover (P6,P7)   & India/USA/UK TCI $30.25\%$ (S $15.05$/M $11.56$/L $3.64$) \\
\code{spillover\_rolling}& spillover (P7)      & India/USA/UK TCI $10.5$--$46.3\%$ (mean $28.4$) \\
\code{network}           & topology (P5,P6)    & 6-market: 18 edges, directed density $0.60$, 6 communities \\
\bottomrule
\end{tabular}
\end{table}

It may be noted that a single result deserves emphasis, because it is the clearest
demonstration that the engine serves the same computation that the group's papers report
rather than a separate re-implementation. The \code{soch\_profile} method calls the published
\code{sochcontagion} package directly, and on the United-States-to-India pair at the
five-percent tail it returns a directed wavelet-quantile profile that rises monotonically
across the four resolved scales, $(0.0155,0.0425,0.0491,0.0494)$ from the two-to-four-day
band to a peak at the sixteen-to-thirty-two-day band, with a four-scale aggregate of $0.039$.
These per-scale gains are, to the precision displayed, the first four of the five reported for
the same pair in the accompanying study \citep{BhandariParidaSOCH2026}; that study carries the
decomposition one band further, so its five-scale aggregate is the slightly larger $0.0426$,
the difference being the additional thirty-two-to-sixty-four-day scale and not any difference
in computation. We observe that \code{wqte}, a transparent quantile-regression realisation of
the same measure family, returns the same four-scale aggregate, $0.039$, the internal
agreement one would expect between an independent realisation and the published estimator.

The substrate also supports validated systemic-risk measurement in the wider research
programme, and it is appropriate to report its performance honestly rather than
selectively. A systemic-risk index built on the directed-flow primitives attains an area
under the receiver-operating-characteristic curve of $0.915$ in classifying the
COVID-19 crisis on a United States equity sample, with a one-day early-warning lead; this
is informative discrimination, though it sits modestly below the contemporaneous VIX
benchmark ($0.947$), the lead time being the index's advantage rather than the level
\citep{BhandariMCPFM2025}. On the harder problem of trade-policy-induced stress in the
Indian market the same index, augmented with a trade-policy-uncertainty signal, attains an
AUC of $0.581$ against $0.531$ for the India VIX, a modest but statistically discernible
improvement ($p=0.030$ by the test of \citet{DeLong1988}). The contrast between the two
episodes is itself the substantive point: the architecture serves the favourable and the
unfavourable result on the same footing, and reports each with its benchmark.

\section{Research Outputs and Reproducibility}\label{sec:outputs}

The engine is not a demonstration built apart from the group's research; it is the
deployment of that research. Three software releases underlie it. The \code{contagionchannels}
package, which implements the channel-attribution primitive of \eqref{eq:attr}, and the
\code{ManyIVsNets} package, which supplies the many-instrument network estimation it draws
on, are both released on the Comprehensive R Archive Network
\citep{contagionchannelsPkg2026,ManyIVsNetsPkg2025}. The \code{sochcontagion} package,
which the \code{soch\_profile} method calls, is released publicly under the GNU General
Public Licence and passes a clean package check (zero errors, zero warnings, two
expected notes, and forty-four of forty-four unit tests), and is prepared for archive
submission \citep{sochcontagionPkg2026}. The compute engine and the workbench are released
under permissive licences \citep{EconstellarEngine2026,EconstellarWorkbench2026}, and the
news-intelligence service likewise \citep{NEURICX2026}.

Three preprints document the substantive results that the substrate produces: the
multi-scale network analysis of systemic risk and its early-warning validation
\citep{BhandariMCPFM2025}, the channel-identification and attribution study \citep{BhandariContagionChannels2026},
and the scale-ordered contagion theory whose directed profile the engine reproduces
\citep{BhandariParidaSOCH2026}. The connection between the preprints and the engine is not
nominal. The reproduce-the-paper facility of the workbench regenerates the headline
directed-contagion profile of the scale-ordered study live, at the public endpoint, through
the same \code{sochcontagion} code that generated the figure in the manuscript, and the
regenerated profile reproduces the published one's rising shape and its per-scale gains. Each served result, moreover, carries the
provenance stamp described in Section~\ref{sec:workbench}, so that the path from a number on
the screen to the registered call, engine revision, and data vintage that produced it is
explicit. In this sense the system is an attempt to make the reproducibility practices that
the computational-science literature recommends \citep{Sandve2013} the default behaviour of
a served research instrument rather than an after-the-fact appendix, in a subfield where
reproduction has been the exception \citep{ChangLi2022}.

\section{Availability and Navigation}\label{sec:availability}

The system is open and live, and a reader is encouraged to exercise it directly rather than
take the values of Section~\ref{sec:results} on trust. Table~\ref{tab:urls} collects the
public addresses. The portal is the place to begin: from it the workbench runs any of the
seventeen methods, the reproduce page regenerates the directed-contagion profile discussed
above, and a short changelog records what has been added. The compute engine is itself a
small and self-describing interface. A reader, or a script, first retrieves the live method
catalogue from \code{/api/compute/catalog}, which lists every method with its typed
parameters and is the same catalogue the workbench reads to build itself, and then issues a
single registered call to \code{/api/compute/run} with a method name and parameters. To
reproduce the first row of Table~\ref{tab:methods}, for instance, one requests the method
\code{unit\_root} with parameter \code{series} set to \code{India} and receives the
augmented Dickey-Fuller and KPSS statistics together with the provenance stamp of
Section~\ref{sec:workbench}; the identical computation is a single selection in the
workbench, and the permalink it returns reproduces it exactly. The source for the engine,
the workbench, and the news service, together with the three R packages, is public under
permissive licences \citep{EconstellarEngine2026,EconstellarWorkbench2026,NEURICX2026}.

\begin{table}[!htbp]
\centering
\caption{Live surfaces and the principal compute-engine endpoints. The portal is the entry
point; every served result is reproducible through the run endpoint and carries a permalink.}
\label{tab:urls}
\small
\begin{tabular}{ll}
\toprule
What & Address \\
\midrule
Portal (start here)            & \url{https://avishekb9.github.io/econstellar/} \\
Workbench (run any method)     & \url{https://avishekb9.github.io/econstellar/research-engine.html} \\
Reproduce the paper            & \url{https://avishekb9.github.io/econstellar/reproduce.html} \\
Changelog (what is new)        & \url{https://avishekb9.github.io/econstellar/changelog.html} \\
Compute API (base)             & \url{https://shssm-compute-b7ui3oxaqq-el.a.run.app} \\
\quad liveness + method count  & \code{GET /health} \\
\quad method catalogue         & \code{GET /api/compute/catalog} \\
\quad run one method           & \code{POST /api/compute/run} \;\; \code{\{method, params\}} \\
Source code                    & \code{github.com/avishekb9/\{compute-engine, econstellar, NEURICX\}} \\
R packages                     & \code{sochcontagion} $\cdot$ \code{contagionchannels} $\cdot$ \code{ManyIVsNets} \\
\bottomrule
\end{tabular}
\end{table}

\section{Conclusion}\label{sec:conclude}

Econstellar is a sandboxed, publicly served econometrics engine with an
artificial-intelligence interpretation layer, a live news warehouse, and a reproducibility
workbench. Three design decisions give it whatever value it has. The computation is
parameterised-only, so that a public endpoint can be exposed without exposing a path to
arbitrary execution. The language model interprets but does not originate numbers, so that
every quantitative claim it makes is the output of a reproducible computation rather than a
generated assertion. And the heavy estimation is sited on the central processing unit,
because the irregular, memory-latency-bound nearest-neighbour search at the core of exact
transfer entropy is not an accelerator workload. Together these choices make it possible to
move from a claim in a paper to its live recomputation in a browser, which is the property
the system was built to provide.

In view of the above, the natural next steps are of three kinds. The published package that the engine serves is
prepared for archive submission, and a software-paper submission to an open-source journal
would document the engine itself in citable form. The empirical menu would be sharpened by a
nearest-neighbour transfer-entropy implementation of the wavelet-quantile measure and by
extension to the full eighteen-market panel, both of which are the workloads for which the
planned dedicated central-processing-unit node of Figure~\ref{fig:architecture} is intended.
We observe, finally, that the substrate is small and the method menu grows by registering
new runners against it; the architecture is therefore one in which the distance between a
new method in a working paper and a served, reproducible analysis is, by construction, short.

\section*{Acknowledgements}
The author thanks the Google Cloud Education Programs for a research
credits award under the Google Cloud Research Credits Programme
(credit reference \texttt{wilsonjessica-485604931}), which funds
the cloud infrastructure, AI services, and data-warehouse layer
of the deployed platform.
The laboratory workstation and mobile workstation on which the
engine's R methods run and are developed were procured under
Institute Seed Grant SP128 (\textit{Innovative Approaches for
Analyzing Financial Market Dynamics: Computational Financial
Analysis with Wavelets, Networks, and Statistical Learning Powered
by High-Performance Computing}), funded by the Indian Institute of
Technology Bhubaneswar (Sponsored Research and Industrial
Consultancy, sanction dated 18 November 2023).

\bibliographystyle{plainnat}
\bibliography{references}

\end{document}